\documentclass[10pt,conference]{IEEEtran}
\IEEEoverridecommandlockouts

\usepackage{algorithmic}
\usepackage{graphicx}
\usepackage{textcomp}
\usepackage{tcolorbox}
\usepackage{pifont}
\usepackage{xcolor}
\usepackage{soul}
\usepackage{booktabs}
\usepackage{multirow}
\usepackage{float}
\usepackage{url}
\usepackage{graphicx}
\usepackage{microtype}
\usepackage{tcolorbox}
\usepackage{enumitem}
\usepackage{seqsplit}
\usepackage[utf8]{inputenc}
\usepackage{newunicodechar,graphicx}

\usepackage[noadjust]{cite}
\usepackage{newunicodechar,graphicx}
\usepackage[hidelinks,bookmarks=true]{hyperref}
\usepackage[switch]{lineno}
\usepackage{enumitem}
\usepackage{seqsplit}
\usepackage{pifont}

\setlist[itemize,enumerate]{noitemsep, topsep=0pt, leftmargin=1.0em}
\setlist{nolistsep}

\hypersetup{
	pdfborder          = {0 0 0},
	breaklinks         = true,
	bookmarksnumbered  = true,
	pdfsubject         = {software engineering},
	pdfkeywords        = {Program Comprehension, Unit Test, Test Case, Assertion, Natural Language Processing, Readability, JUnit, Software Quality},
	pdftitle           = {An Exploratory Study on the Usage and Readability of Messages Within Assertion Methods of Test Cases},
	pdfauthor          = {Taryn Takebayashi,
Anthony Peruma, Mohamed Wiem Mkaouer, Christian D. Newman}
}
\DeclareRobustCommand{\okina}{%
  \raisebox{\dimexpr\fontcharht\font`A-\height}{%
    \scalebox{0.8}{`}%
  }%
}
\newunicodechar{ʻ}{\okina}

\usepackage{listings}

\definecolor{javared}{rgb}{0.6,0,0} 
\definecolor{javagreen}{rgb}{0.25,0.5,0.35} 
\definecolor{javapurple}{rgb}{0.5,0,0.35} 
\definecolor{javadocblue}{rgb}{0.25,0.35,0.75} 
 
\newcommand{\RQA}{\textbf{RQ1}: How often do developers use messages in their assertion methods, and in what ways are these messages related?}

\newcommand{\RQB}{\textbf{RQ2}: To what extent do developers provide easily comprehensible messages in assertion methods?}

\begin{document}

\title{\huge An Exploratory Study on the Usage and Readability of Messages Within Assertion Methods of Test Cases}


\makeatletter
\newcommand{\linebreakand}{%
  \end{@IEEEauthorhalign}
  \hfill\mbox{}\par
  \mbox{}\hfill\begin{@IEEEauthorhalign}
}
\makeatother

\author{
\IEEEauthorblockN{Taryn Takebayashi}
\IEEEauthorblockA{\textit{Information and Computer Sciences Department} \\
\textit{University of Hawaiʻi at Mānoa}\\
Honolulu, Hawaiʻi, USA \\
tarynet@hawaii.edu}
\and
\IEEEauthorblockN{Anthony Peruma}
\IEEEauthorblockA{\textit{Information and Computer Sciences Department} \\
\textit{University of Hawaiʻi at Mānoa}\\
Honolulu, Hawaiʻi, USA \\
peruma@hawaii.edu}
\linebreakand
\IEEEauthorblockN{Mohamed Wiem Mkaouer}
\IEEEauthorblockA{\textit{Department of Software Engineering} \\
\textit{Rochester Institute of Technology}\\
Rochester, New York, USA \\
mwmvse@rit.edu}
\and
\IEEEauthorblockN{Christian D. Newman}
\IEEEauthorblockA{\textit{Department of Software Engineering} \\
\textit{Rochester Institute of Technology}\\
Rochester, New York, USA \\
cnewman@se.rit.edu}
}


\maketitle

\begin{abstract}
Unit testing is a vital part of the software development process and involves developers writing code to verify or assert production code. Furthermore, to help comprehend the test case and troubleshoot issues, developers have the option to provide a message that explains the reason for the assertion failure. In this exploratory empirical study, we examine the characteristics of assertion messages contained in the test methods in 20 open-source Java systems. Our findings show that while developers rarely utilize the option of supplying a message, those who do, either compose it of only string literals, identifiers, or a combination of both types. Using standard English readability measuring techniques, we observe that a beginner's knowledge of English is required to understand messages containing only identifiers, while a 4\textsuperscript{th}-grade education level is required to understand messages composed of string literals. We also discuss shortcomings with using such readability measuring techniques and common anti-patterns in assert message construction. We envision our results incorporated into code quality tools that appraise the understandability of assertion messages.  
\end{abstract}


\section{Introduction}
Unit tests are essential to ensuring a software system's quality \cite{pressman2014software}. Unit tests involve writing code in the form of test methods (i.e., test cases) to verify the functionality of the source (i.e., production) code. A critical part of a test method are assertion statements \cite{Bowes2017WETSoM}. These are methods that developers utilize to verify the output of a production method, under test, against a predefined value. The result of an assertion method determines if the test case passes or fails. Further, to help developers with troubleshooting failing test cases, developers have the option to provide a message explaining the reason for the failure.

Similar to production code, developers must adhere to best practices when writing unit tests so as not to impede maintenance activities \cite{Welker1997}. To this extent, current research on test code quality has predominantly focused on test smells \cite{Aljedaani2021EASE}, identifier naming \cite{Peruma2021ICPC,Daka2017ISSTA}, including the impact of rename recommendation models \cite{Lin2019SCAM}, and test case readability \cite{Setiani2020Access, Daka2015FSE, Grano2018ICPC}. While these research works do indeed focus on the quality of test code, studies utilizing or proposing readability models do not consider the explanation message of an assertion method. At most, these code readability models consider the presence of assertions in the test case \cite{Setiani2020Access, Daka2015FSE} or are not specialized for test code \cite{Scalabrino2016ICPC}. Furthermore, most test readability studies focus on the readability of automatically generated test code or comparing the readability of developer-written test cases against auto-generated code \cite{Winkler2022SANER}. Since developers are free to have multiple assertions within a test case and are not restricted in how they craft the message, having an understandable message is imperative to comprehending both test case behavior and troubleshooting test failures. For example, when comparing the messages in Listing \ref{Listing:example1} and Listing \ref{Listing:example2}, we see that the message in Listing \ref{Listing:example1} is readable and descriptive, clearly stating the reason for the failure of the test case, which helps the developer with troubleshooting. However, the message in Listing \ref{Listing:example2} is a set of numbers (enclosed within a string literal), and while it might make sense (i.e., understandable) to the author of the test case, external or new team developers would find it challenging to understand the meaning of the message.

Since little is known about the extent to which developers rely on the documentation feature of assertion methods, this exploratory study forms the first step to more in-depth studies of assertion messages. That, for instance, investigating the benefits developers gain from these messages and the developer's mental model for crafting and comprehending messages.

\begin{lstlisting}[caption=An example of an assertion method with a textual message explaining the test case's failure \cite{intro-example1}., label=Listing:example1, firstnumber = last, escapeinside={(*@}{@*)}]
@Test
public void testQuery() {
 .
 .
 Iterator<QueryKeyResult> results;
 results = getAdminClient().streamingOps.queryKeys(0, testStoreName, queryKeys.iterator());
 assertTrue("Results should not be empty", results.hasNext());
 .
 .
}
\end{lstlisting}

\begin{lstlisting}[caption=An example of an assertion method using digits (contained within a string literal) to explain the test case's failure \cite{intro-example2}., label=Listing:example2, firstnumber = last, escapeinside={(*@}{@*)}]
@Test
void testIsEmpty() {
 .
 .
 assertTrue("(0,0)", new Range<>(Integer.class, 0, false, 0, false).isEmpty());
 .
 .
}
\end{lstlisting}

\subsection{Goal \& Research Questions}
The goal of this study is to explore the use of explanation messages in assertion methods contained within test methods. Thus, we study \textit{the extent to which developers utilize these messages and the characteristics of how the messages are structured to help developers efficiently troubleshoot test case failures.} We envision findings from our study supporting the development of tools and techniques for appraising the quality of assertion messages and supporting auto-generation of test cases with comprehension-friendly messages. Hence, we answer the following research questions (RQs): 

\vspace{1.3mm}
\noindent\textbf{\RQA} This question informs us of the extent to which developers utilize an explanation message in an assertion method contained within test methods. Furthermore, through this RQ, we group messages into high-level categories based on the types of tokens present in the message to highlight how developers compose the message.

\vspace{1mm}
\noindent\textbf{\RQB} In this RQ, we utilize existing techniques to measure the readability of different categories of assertion messages and supplement the findings with qualitative examples. At a high level, the results provide developers with guidance on how they should craft their messages.

\subsection{Contribution}
The main contributions from this work are as follows:
\begin{itemize}
    \item Our results represent a preliminary yet significant step toward a deeper awareness of assertion messages. Through our discussion, we pave the way for subsequent research that will enhance our knowledge of high-quality comprehensible assertion messages, especially in the realm of automated code quality tools that integrate into the development workflow.
    \item A dataset of assertion messages, their categorization, and linguistic properties for extension and replication purposes.
\end{itemize}

\subsection{Paper Structure}
This paper is organized as follows: In Section \ref{Section:related}, as part of our related work, we report on studies that examine the readability of unit test code. Section \ref{Section:experiment_design} provides details about our investigation methodology, while Section \ref{Section:experiment_results} answers our research questions by reporting on the results of our experiments. Section \ref{Section:discussion} provides an in-depth discussion and takeaways from our findings. Finally, Section \ref{Section:threats} reports on the threats to the validity of our study before our conclusion and future work in Section \ref{Section:conclusion}.



\section{Related Work}
\label{Section:related}
While there are multiple studies on code readability models (e.g., \cite{Buse2010TSE,Scalabrino2016ICPC}), empirical code readability studies involving project/code repositories and developers/students (e.g., \cite{Fakhoury2019ICPC,Santos2018ICPC,Oliveira2020ICSME}), we limit our reporting of related work to only studies examining test code's readability. These studies include generating/evaluating test code identifier names, specifically test method names, and work that proposes test readability models or utilizes readability models on test code.

Wu and Clause \cite{Wu2020JSS} utilize a set of test method construction patterns (i.e., action-predicate-scenario) to identify non-descriptive test method names and provide developers with information for a more descriptive name. The authors conduct an empirical study and show that their technique yields a 95\% true-positive rate. In another study \cite{Wu2022TOSEM}, the authors compare a test method name with its siblings and determine that its name is often based on what makes the test unique from its siblings. The authors propose an automated technique that extracts unique attributes of a test, and an evaluation by subject matter experts shows that these details are helpful for test method name generation. Zhang et al. \cite{Zhang2015ASE} parse test method names using natural language processing techniques to produce test method templates automatically. 
The action phrase and the predicate phrase from the name of a test method are used by the authors in their strategy, which relies on elements of English grammar to carry out the parsing. The authors report an accuracy of 80\% of their proposed technique. In a later work \cite{Zhang2016ASE}, the authors propose a natural language processing-based technique for generating descriptive test method names based on the method's body. In their strategy, the authors examine the test method's assertions to determine the action, anticipated result, and the situation being tested. In their approach, Daka et al. \cite{Daka2017ISSTA} suggest a method for generating concise, descriptive test method names based on API-level coverage targets, and validate their method by interviewing 47 students. Lin et al. \cite{Lin2019SCAM} examine the quality of identifiers in test suites and compare them to production identifiers. The results show that identifiers in test suites are of low quality, especially identifiers in automatically generated test code. In an empirical study by Peruma et al. \cite{Peruma2021ICPC}, the authors examine the evolution of test method names by examining how the part-of-speech tags change during rename operations. The authors show that test method names typically have a structure that differs from production method names. The authors also show some common pattern structures developers utilize for naming test methods.

Daka et al. \cite{Daka2015FSE}, propose a readability model for test code that utilizes multiple features such as identifier length, numbers, loops, asserts, etc. The authors show that their approach outperforms other readability modes on code snippets. However, although the author's model includes assertions, the authors do not consider the understandability of the assert method message. In an extension of their work \cite{daka2015generating}, the authors utilize the model of test readability and additional optimization techniques for test generation for the Guava library and show that subject matter expert prefer their readability-optimized test cases. Using a readability model, an empirical study by Granoet al. \cite{Grano2018ICPC} shows that test code is less readable than the  production code they test. The authors also state that developer-written test cases are easier to read than automatically generated test cases. 

\subsection{Summary}
While there are studies on test code readability, these studies do not evaluate the quality of the explanation message in the assert method; at most, they limit their analysis to the presence/absence of an assert method in a test case. In short, we lack the understanding of knowing if and how developers use the documentation property of assertion methods.

\section{Experiment Design}
\label{Section:experiment_design}
In this section, we discuss the methodology of our study. Figure \ref{Figure:diagram_experiment} depicts an outline of the activities in our experiment methodology, which we describe below. Furthermore, our dataset is available for replication/extension purposes at \cite{website}.

\begin{figure}
 	\centering
 	\includegraphics[trim=0cm 0cm 0cm 0cm, width=1\linewidth]{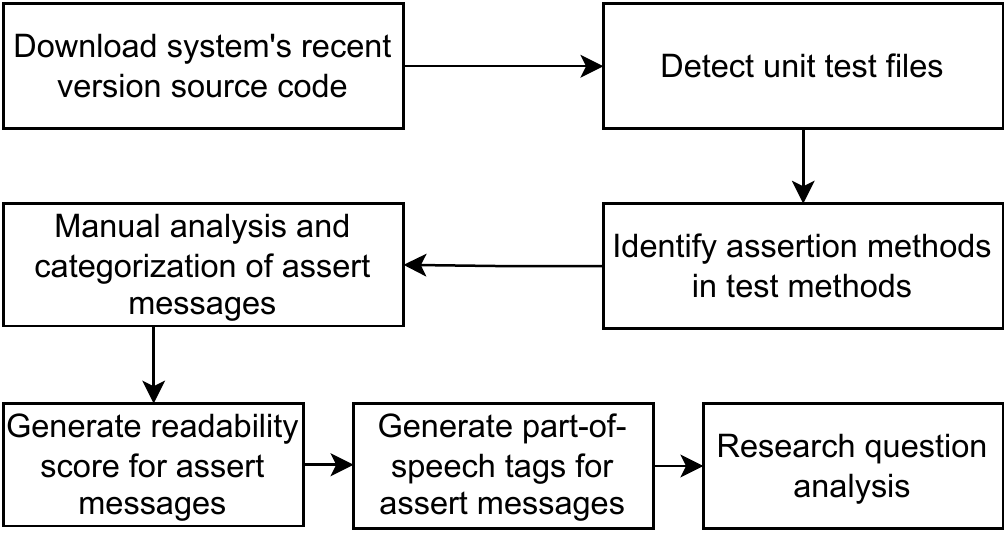}
 	\caption{Overview of our experiment design.}
 	\label{Figure:diagram_experiment}
\end{figure}

\vspace{1mm}
\noindent\textbf{Source Projects.} This study analyzes the most recent release/version (at the time of this study) of 20 open-source Java systems that have their unit tests written using the JUnit 4 testing framework\footnote{\url{https://junit.org/junit4}}. We selected these systems as they were used in prior unit test code-related studies \cite{coro2020jtec, Wu2022TOSEM, Peruma2021ICPC}.

\vspace{1mm}
\noindent\textbf{Test File Detection \& Parsing.} We utilized the JetBrains Software Development Kit (SDK) \footnote{\url{https://www.jetbrains.com/help/idea/sdk.html}}, specifically the IntelliJ Platform SDK\footnote{\url{https://plugins.jetbrains.com/docs/intellij/about.html}}, to detect test files in the projects and parse these files. The SDK provides a rich API, including utility methods, to help us with our automation tasks. For each project, we first obtained the list of JUnit test files using the \texttt{JUnitUtil.getTestClass()}\footnote{\url{https://github.com/JetBrains/intellij-community/blob/master/java/execution/impl/src/com/intellij/execution/junit/JUnitUtil.java}} API. This method, internally, utilizes multiple heuristics, such as evaluating annotations, to determine if the class is a valid test file. Next, we utilize \texttt{JavaRecursiveElementVisitor}\footnote{\url{https://github.com/JetBrains/intellij-community/blob/idea/223.7571.182/java/java-psi-api/src/com/intellij/psi/JavaRecursiveElementVisitor.java}} to access each method declaration in the test class and obtain the list of method call expressions within the method declaration. Next, we evaluate each method call to determine if the called method is a JUnit assertion method\footnote{\url{https://junit.org/junit4/javadoc/latest/org/junit/Assert.html}}. An advantage of using the IntelliJ Platform SDK is that we can check the class the called method belongs to and the data type of the method's parameters. Using these features, we do not have to only depend on the method's name to determine if it is a JUnit assert, hence, ensuring that we only include valid JUnit assertion methods in our experiments. For all detected assertions, we extract the assert method and other metadata, such as the containing method, class, and line number, and if the containing method is a test method.



\vspace{1mm}
\noindent\textbf{Manual Message Categorization.} To better understand the types of messages developers provide in assertion methods, we manually reviewed each message in the dataset. We assign the message to one of the three below categories as part of the review \textit{(note: in the below examples, we highlight, in yellow, the assertion message)}:
\begin{itemize}
    \item Identifier - The message is the value of an identifier, such as a method call, a constant, or variable/attribute value e.g.,  \texttt{assertEquals(\hl{r.getLowerBound()}, 1.0, EPSILON);}
    \item Text - The message is a string literal (i.e., a sequence of characters enclosed in double quotation marks)\newline e.g.,  \texttt{assertNull(\hl{"interpolate"}, name);}
    \item Combination - The message is a combination of string literals and identifiers appended together\newline e.g.,  \texttt{assertFalse(\hl{``too big at '' + count}, count > 100);}
\end{itemize}

\vspace{1mm}
\noindent\textbf{Readability Score Generation.} We leverage multiple well-established English text readability metrics to determine the understandability scores for the assertion messages. These metrics have been utilized in prior work as-is or in conjunction with other metrics/measurements to assist with code readability (e.g., comprehending code comments) \cite{Scalabrino2016ICPC, Khamis2010NLPIS, Eleyan2020Information}. We utilize the following readability tests:
\begin{itemize}
    \item Flesch Reading Ease \cite{Flesch1948} - The higher the score, the easier it is to read; text with a high score usually has short sentences
    \item Flesch-Kincaid Grade Level \cite{Kincaid1975} - Calculates the minimum U.S. education grade required to understand the text 
\end{itemize}

Since the message text can contain an identifier name with more than one term, such names must be split into their constituent terms. For example, the method name \texttt{getLowerBound()}, should be split into the terms `get', `Lower', and `Bound'. Similar to prior work on identifier names \cite{Peruma2021ICPC,Peruma2020JSSRenames,Peruma2018IWoR}, we split such names using the Ronin splitter algorithm implemented in the Python package Spiral \cite{huckaspiral}. This pre-processing ensures more accurate readability scores. Furthermore, we also extract measures around the structure of the message, such as the number of words, sentences, polysyllabic words, etc., using the Python Textstat\footnote{\url{https://github.com/textstat/textstat}} library.

\vspace{1mm}
\noindent\textbf{Part-of-Speech Generation.} To determine the part-of-speech tags associated with the words in an assertion message, we leveraged the default Perceptron tagger available in Python's NLTK package \cite{Bird2009NLTK}.

\vspace{1mm}
\noindent\textbf{Research Question Analysis.} We use a mixed-methods strategy to examine the mined/extracted data. This approach to analyzing quantitative and qualitative data allows us to provide representative samples from the dataset to complement our findings. Details of our analysis are detailed in Section \ref{Section:experiment_results}.

\section{Experiment Results}
\label{Section:experiment_results}
In this section, we report on the findings of our experiments by answering our RQs. The first RQ examines the volume and types of assertion methods containing an explanation message. The second RQ investigates the readability of these messages. Due to space constraints, specific tables in the RQs show only the most frequently occurring instances; the complete set is available at: \cite{website}.

\subsection*{\RQA}
In the first part of this RQ, we examine the volume of assertion methods containing an explanation message. 

Our analysis of 20 projects yields 31,063 methods that contain one or more JUnit assertions statements. However, out of this, only 1,714 (or approximately 5.52\%) methods have one or more assertions with an explanation message. Furthermore, when we look at the individual assertion statements, we encounter a total of 86,321 JUnit assert methods. From this, only 3,872 (or approximately 4.49\%) assertion methods contained an explanation message. It should be noted that while every project in the dataset has assertions, only 11 projects have assertions with an explanation message, with a median of 115 assertion methods. Table \ref{Table:stat_summary} shows a statistical summary of the occurrence of assertion methods with and without an explanation message in each project within our dataset.


\begin{table}
\centering
\caption{Statistical summary on the occurrence of assertion methods in each project within the dataset.}\vspace{-3mm}
\label{Table:stat_summary}
\begin{tabular}{@{}crrrrr@{}}
\toprule
\textbf{Min.}          & \multicolumn{1}{c}{\textbf{1st Qu.}} & \multicolumn{1}{c}{\textbf{Median}} & \multicolumn{1}{c}{\textbf{Mean}} & \multicolumn{1}{c}{\textbf{3rd Qu.}} & \multicolumn{1}{c}{\textbf{Max.}} \\ \midrule
\multicolumn{6}{c}{\textit{Assertion methods without an explanation message}}                                                                                                                                           \\
\multicolumn{1}{r}{1.0} & 58.0                                  & 270.0                               & 4,122.0                                & 2,826.0                            & 42,573.0                                        \\ \midrule
\multicolumn{6}{c}{\textit{Assertion methods with an explanation message}}                                                                                                                                              \\
\multicolumn{1}{r}{2.0}  & 14.5                                   & 115.0                                  & 352.0                                  & 523.5                             & 1,335.0                                        \\ \bottomrule
\end{tabular}
\end{table}

Our subsequent analysis examines the type of assertion method frequently containing an explanation message. As shown in Table \ref{Table:assertType}, the majority of the explanation messages are contained within an \texttt{assertEquals} (40.34\%), followed by the \texttt{assertTrue} method (31.02\%). Next, we examine the common unigrams and bigrams developers utilize when constructing a message. While we did notice that developers utilize similar terms for \texttt{assertEquals} and \texttt{assertTrue}, such as `should', `to\_string', we also did encounter the bigram `does\_not' frequently appearing within \texttt{assertEquals}. Developers use this bigram to state that the expected value does not match the actual value, as in the example:  \texttt{assertEquals(``invocation time does not match'', expected.getInvocationTime(), actual.getInvocationTime())}. Looking at the frequent terms in  \texttt{assertNotNull}, we observe the unigram `null', where developers use this term to show that result of the assertion is a null value when it should not, as in  \texttt{assertNotNull(``style is null'', style)}. Looking at the messages within the \texttt{fail} assertion method, we encounter the terms `error', `failed\_to', and `not\_thrown'. The \texttt{fail} method is typically utilized to fail a test case if an expected exception does not occur. In other words, the success of the test case depends on the occurrence of an exception, as in the example: \texttt{fail(``Error in predicate but not thrown!'')}.


\begin{table}
\centering
\caption{Distribution of the top four assertion method types frequently containing an explanation message.}\vspace{-3mm}
\label{Table:assertType}
\begin{tabular}{@{}lrr@{}}
\toprule
\multicolumn{1}{c}{\textbf{Assertion Type}} & \multicolumn{1}{c}{\textbf{Count}} & \multicolumn{1}{c}{\textbf{Percentage}} \\ \midrule
\texttt{assertEquals}                      & 1,562                              & 40.34\%                                 \\
\texttt{assertTrue}                              & 1,201                              & 31.02\%                                 \\
\texttt{assertFalse}                        & 570                                & 14.72\%                                 \\
\texttt{assertNotNull}                       & 229                                & 5.91\%                                  \\
\textit{Others}                            & 310                                & 8.01\%                                  \\ \midrule
Total                             & 3,872                              & 100\%                                   \\ \bottomrule
\end{tabular}
\end{table}

\begin{table}
\centering
\caption{Distribution of how assert messages are structured.}
\label{Table:Category}
\begin{tabular}{@{}lrr@{}}
\toprule
\multicolumn{1}{c}{\textbf{Category}} & \multicolumn{1}{c}{\textbf{Count}} & \multicolumn{1}{c}{\textbf{Percentage}} \\ \midrule
String Literal            & 3,135 & 80.97\% \\
Combination               & 414   & 10.69\% \\
Identifier                & 323   & 8.34\%  \\ \midrule
\multicolumn{3}{c}{\textit{Combination}}    \\
String + Variable         & 236   & 57.00\% \\
String + Method           & 159   & 38.41\% \\
String + Variable + Method & 13    & 3.14\%  \\
String + Digit            & 6     & 1.45\%  \\ \midrule
\multicolumn{3}{c}{\textit{Identifier}}     \\
Method                    & 229   & 70.90\% \\
Variable                  & 94    & 29.10\% \\ \bottomrule
\end{tabular}
\end{table}

Moving on, based on our categorization of the assertion messages (described in Section \ref{Section:experiment_design}), we observe that the majority of these messages are composed of string literals (3,135 instances or 80.97\%), followed by a combination of string literals and identifiers
(414 instances), and identifiers (323 instances). Furthermore, when looking at the identifier category, we observe that developers prefer using methods over variables, where developers usually utilize the \texttt{toString()} method of an object to generate the message for the assertion, like \texttt{assertEquals(map.toString(), expected, map.size())}. In contrast, the combination category shows developers preferring to use variables over methods in conjunction with string literals, like \texttt{assertEquals("Incorrect id: " + actual, expected, actual)}. Table \ref{Table:Category} shows a breakdown of the distribution of these categories in our dataset.


\begin{tcolorbox}[top=0.5pt,bottom=0.5pt,left=1pt,right=1pt]
\textbf{Summary for RQ1.}
Even though assert methods are an essential part of unit testing, most developers do not make use of the ability to provide a message to explain the failure of a test case. Developers frequently provide messages using either the \texttt{assertEquals}  and \texttt{assertTrue} methods. Depending on the type of assert method used, developers tend to use specific terminology when crafting their message. Finally, we observe three high-level patterns developers utilize to craft messages, with developers frequently using a string of literals for the message.
\end{tcolorbox}

\subsection*{\RQB}

In this RQ, we leverage standard English readability metrics to determine the level of difficulty associated with comprehending assertion messages. Since an assertion message is meant to assist developers in understanding the failure of the assertion, and developers are free to compose this message using a variety of words, it is reasonable to utilize such metrics. Furthermore, we examine how developers structure these messages by examining the part-of-speech tags. To this extent, we utilize the readability models and part-of-speech tagger described in Section \ref{Section:experiment_design}. We complement the generated quantitative data with qualitative examples from the dataset.

\vspace{2mm}
\noindent\textit{Message Readability}
\vspace{1mm}

In this part of the RQ analysis, we report on the readability scores associated with the assert messages in our dataset. The readability of an English sentence indicates whether or not it contains words associated with higher education levels. 

First, we look at messages composed of only string literals (i.e., pure text messages). String literals were the most common message type in this dataset; of the 3,872 total messages, 3,135 (80.97\%) were composed of only string literals. The Flesch reading ease median score for such messages is 77.91, and the mode was 93.81, indicating that most messages ranged from fairly easy to read to plain English \cite{Flesch1979Write}. Furthermore, as per the Flesch-Kincaid grade level score, such text requires a 4\textsuperscript{th}-grade education level to be understood. Looking at the structure of the messages, we observe that these messages have a median and mean of one sentence, a median of five tokens, and less than one polysyllabic word; such traits help with text readability, and understandability \cite{Flesch1948}.

Next, we analyze messages composed of only identifiers (i.e., method calls or variables). Our analysis yields 323 (8.34\%) messages composed of only identifiers with a Flesch reading ease median score of 93.81 and a corresponding grade level of less than 1 (i.e., a beginner's knowledge of English). These metrics, along with a median of five tokens and less than one polysyllabic word, seem to indicate high readability for identifier messages. However, in most cases, the value returned by the identifier does not provide details about why the test fails. For example, in the following assertion \texttt{assertNotNull(msg,c1}, the message will be printed if c1 is null. If the message consisted of the string ``c1 should not be null" that would be an appropriate message. Furthermore, the message should contain c1's type. However, we don't know the contents of the message. 
Therefore, the standard readability indicators used are not good indicators of the readability of identifier messages. 

Moving on, our examination of the 414 (10.69\%) messages composed of a combination of text and identifiers shows a median reading ease of 78.25 and a median reading comprehension of a 4\textsuperscript{th}-grade education level. These types of combination messages tended to contain less than 1 polysyllable. However, due to the presence of identifiers, combination messages tend to have a greater number of tokens (median token count of 11). In addition, the most common types of messages began with a string literal followed by an identifier, such as \texttt{"Expected: " + expected + ", got:" + actual}. These messages have a Flesch reading ease score of -8.73, suggesting they are very difficult to read. Furthermore, we observe combination messages with high Flesch reading ease scores composed of empty strings added to an identifier or very short sentences. For example, in \texttt{assertTrue( assertTrue("" + c, c $>=$ 100)}, an empty string is appended to an identifier and has a Flesch reading ease score of 121.22. Additionally, \texttt{assertTrue("file " + file.getAbsolutePath() + " should exist", file.exists())} is an example where a single word is appended to an identifier and has a reading score of 81.29.

Finally, it should be worth noting that high readability does not inherently indicate high understandability. For example, single-word messages, such as `works', `method', and `found' are easy to read (Flesch reading ease score of 121.22). However, by themselves, these words hold little meaning in the context of test case failure. At the same time, not all single-word messages mean easy readability. Technology/domain terms such as `Serialization' result in negative reading scores (in this case -217.19). Furthermore, the presence of these domain/technology terms in longer messages also yields poor readability scores. For instance, ``Serialized and deserialized value is different'' has a Flesch reading ease score of -27.68. This shows that standard English language readability models, which may work well for general-purpose English prose, may not be an ideal solution for measuring source code readability. 

\vspace{2mm}
\noindent\textit{Message Structure}
\vspace{1mm}

In this part of our RQ analysis, we examine the part-of-speech tags developers utilize to construct messages. We limit our analysis to only messages that we categorize as consisting of only string literals. We did not analyze messages constructed with identifiers as prior work shows that developers utilize a different structure when naming identifiers \cite{Newman2020JSS}, and hence a comparison is not feasible. Table \ref{Table:pos} shows the top two frequently occurring patterns for the first term, the first two terms, and the first three terms in a message. 

As shown in Table \ref{Table:pos}, most messages start with either a  noun (i.e., a singular noun or proper noun-- 25.01\% and 21.5\%, respectively). Typically, in the English language, a sentence starts with a specific subject, which is often a noun, and is followed by a verb \cite{Cutts2020}; this aligns with our observation of assert messages starting with a type of noun. Examining a sample of these messages, we notice that the noun is usually a reference to the entity under test, as in the case of \texttt{(assertFalse("File still exists after deletion", Files.exists(symLinkP, NOFOLLOW\_LINKS)))} where the test checks if a file deletion was processed correctly. Next, looking at the first two terms in a message, we observe that developers typically construct messages starting with an adjective followed by a noun, such as \texttt{"Invalid number"}. Furthermore, an adjective + noun structure can also be described as a complete subject, where an adjective (also known as noun-adjunct) describes specific details that define the subject. A noun-adjunct is a word that is typically a noun but is being used as an adjective \cite{Newman2020JSS}. Likewise, we observe a similar pattern when examining the first three terms in a message, where developers provide a more descriptive text.


Other patterns of interest outside of the top two patterns mentioned in Table \ref{Table:pos} were messages that started with a verb, namely a past participle verb. Similarly to adjectives, past participle verbs often function in modifying or describing a noun. So, seeing assert messages beginning with a past-participle verb is not unusual when compared to regular English sentences. In most cases, the starting verb is mapped to the term `expected', which is utilized in messages about exceptions that should occur in a test case. For example, in the assertion \texttt{fail("Expected IllegalStateException.")}, the term starts with a verb and is followed by a noun that represents the type of exception. In some cases, the antonym `unexpected' was used in a similar fashion, \texttt{assertEquals("Unexpected map size", 2, map.size()))}. Other starting verbs included `Returned', `Interpolated', and `Created' followed by a noun referencing the entity under test, and the modal term `should', \texttt{assertEquals("Returned value should be equal to BigInteger.ONE", BigInteger.ONE,value)}.





\begin{table}
\centering
\caption{Common part-of-speech patterns for text-based messages in an assertion method.}
\label{Table:pos}
\begin{tabular}{@{}lrr@{}}
\toprule
\multicolumn{1}{c}{\textbf{Part-of-Speech Pattern}} & \multicolumn{1}{c}{\textbf{Count}} & \multicolumn{1}{c}{\textbf{Percentage}} \\ \midrule
\multicolumn{3}{c}{\textit{First term in a message}}                                                                               \\
Noun (singular)                                                & 784                              & 25.01\%                                 \\
Proper noun (singular)                                     & 674                                & 21.50\%                                 \\
\textit{Others}                                     & 1,677                              & 53.49\%                                 \\ \midrule
\multicolumn{3}{c}{\textit{First two terms in a message}}                                                                          \\
Adjective, Noun (singular)                                  & 332                                & 11.32\%                                 \\
Proper noun (singular), Noun (singular)                   & 219                                & 7.46\%                                  \\
\textit{Others}                                     & 2,383                              & 81.22\%                                 \\ \midrule
\multicolumn{3}{c}{\textit{First three terms in a message}}                                                                        \\
Adjective, Noun (singular), Noun (singular)                    & 81                                & 3.14\%                                 \\
Adjective, Verb, Noun (plural)                        & 73                                & 2.83\%                                  \\
\textit{Others}                                     & 2,429                              & 94.04\%                                 \\ \bottomrule
\end{tabular}
\end{table}

\begin{tcolorbox}[top=0.5pt,bottom=0.5pt,left=1pt,right=1pt]
\textbf{Summary for RQ2.}

Existing English language readability techniques indicate that assertion messages composed of only identifiers are easier to read and require a beginner's knowledge of English. Messages composed of only string literals require a 4\textsuperscript{th}-grade education level to understand and tend to follow a similar grammatical pattern to English sentences, most often beginning with a subject or complete subject (adjective/modifier + noun).
\end{tcolorbox}

\section{Discussion \& Takeaways}
\label{Section:discussion}
Assertion methods are an essential part of a test case. Furthermore, to assist developers with understanding the failure of a test case, these assertion methods provide developers the opportunity to provide an optional message explaining the failure. Therefore, these messages must be readable and understandable to ensure developer productivity and system quality during troubleshooting and onboarding activities. In this study, we explore the extent to which developers utilize messages in assertion methods and the characteristics of these messages concerning readability. Our findings show that although the volume of assertion methods containing messages is low, the messages that do exist can be grouped into one of three high-level categories. While our findings extend the knowledge of code readability, there are avenues for further research, including the evolution of messages and how developers comprehend these messages. Below, we discuss how the findings from our RQs support the community through a series of takeaways. 

\vspace{1mm}
\noindent\textbf{Takeaway 1 - \textit{Encourage the use of messages in assertion methods.}}
Since assertion methods are optional, developers rarely use the opportunity to provide a message. However, having multiple non-documented assertions impedes troubleshooting test failure and is a known, commonly occurring test smell (i.e., Assertion Roulette) \cite{meszaros2007xunit,panichella2022test, bavota2015test}. To this extent, academia should instill in students the importance of using messages in assertions. Additionally, static analysis and test suite generation tools should incorporate the importance of using messages in asserts. Finally, based on our data and experiences during this study, the research community needs to focus on how we can support developers in creating effective, concise assertion messages. In other words, are there ways we can generate messages? Are there ways of phrasing messages that improve comprehension for different types of asserts?

\vspace{1mm}
\noindent\textbf{Takeaway 2 - \textit{Extend the applicability of readability quality metrics.}}
Our RQ2 findings show that English readability techniques consider messages composed of string literals as descriptive and follow a fairly common grammatical structure similar to general English prose. The heuristics we find in this study can help code readability models, which currently do not directly support messages within assertion methods. For example, they should include heuristics to warn developers against using messages composed of only digits, which is also a test smell (i.e., Magic Number Test) \cite{Peruma2019CASCON}.

\vspace{1mm}
\noindent\textbf{Takeaway 3 - \textit{Establishment of a catalog of anti-patterns in assert message usage.}}
In our manual analysis of the data, we observe instances of assert methods having messages that, while they seem readable, are inclined to misinterpretation due to the terminology developers use in constructing the message. Similar to linguistic anti-patterns \cite{Arnaoudova2016EMSE}, these mistakes or misconceptions can hinder communication or understanding. Below we provide a summary of the anti-patterns we identified. 
\begin{itemize}
    \item \textit{Message does not say what is expected.} In this anti-pattern, the assert message written by the developer does not provide information on why the test case fails. For example, in \texttt{\seqsplit{assertEquals("values.size()", 3, content.size())}}, the developer prints the string ``values.size()'', instead of a message stating why the equality test is failing. Most likely, this specific case might have been a mistake by the developer (maybe when debugging); instead of using a method call, the developer enclosed an identifier within quotes.  
      
    \item \textit{Message is misleading.} In this anti-pattern, the meaning of the message does not align with the intention of the test. For example, in \texttt{\seqsplit{assertTrue("available", buffer.available())}}, the message ``available'' is shown if `buffer.available()' returns false. Ideally, the message should be ``buffer is not available''.
    
    \item \textit{Message is too short.} Understandably, this anti-pattern can be subjective as the ideal message length differs by developer. However, a message comprising of a single word may not provide enough context to understand the intended meaning. For example, in \texttt{\seqsplit{assertNull("interpolate", name)}}, the message ``interpolate'' does not provide adequate details about the test's failure. Further, using abbreviations and acronyms in/as messages should be discouraged as prior research on identifier naming shows that developers have difficulty comprehending short names, including those composed of abbreviations and acronyms \cite{Hofmeister2017SANER, Schankin2018ICPC}.
\end{itemize}
While these observations are interesting and further support the fact that the quality of assert messages contributes to the overall quality and understandability of the code, additional studies are required to create a more formal and exhaustive set of assert message anti-patterns.

\vspace{1mm}
\noindent\textbf{Takeaway 4 - \textit{Construction of natural language processing tools specific to assertion message analysis.}}
Our use of standard English readability metrics is an initial step in research in this field. However, prior research has shown the importance of using specialized natural language processing tools in analyzing software artifacts \cite{Newman2021TSE, Binkley:2018}. For instance, messages containing a combination of string literals and identifiers require preprocessing that is not straightforward, and require the use of specialized tools/packages. Additionally, it is essential to understand the context around its use to verify if the message accurately reflects its intended behavior, which can be inferred by analyzing the surrounding code \cite{Newman2020JSS}. Finally, even though the English readability metrics employed in our study are utilized in prior research, additional readability models exist that incorporate conflicting formulas \cite{Zhou2017}, further highlighting the need for a specialized model/tool.  

\section{Threats To Validity}
\label{Section:threats}
Even though our analysis is limited to systems utilizing JUnit 4, other testing frameworks provide similar assertion functionality. Additionally, these open-source systems are utilized in published software engineering research studies. Nevertheless, there is a threat that our findings may not represent industry/closed-source systems. However, as this is an exploratory study, our work provides a starting point for researchers to conduct comparison studies against other systems. While we leverage well-known text readability metrics, these models were originally constructed to analyze English prose. That said, these same metrics have been utilized in prior code readability studies. Our discovery of anti-patterns in how developers craft assert messages is due to manual analysis. Hence, there does exist the possibility of other types of anti-patterns that we might have missed. However, as stated before, this is an exploratory study, and our findings serve as a starting point in understanding the ineffective ways developers construct assert messages. 

\section{Conclusion \& Future Work}
\label{Section:conclusion}
Assertion methods are a crucial element in test cases, which developers utilize to make assertions about the state of the system under test. To help debug test failures and communicate the test's purpose, developers can provide an optional explanation message. Developers are free to compose these messages in any manner they see fit. This study represents an initial foray into the area of assert messages and begins to describe if and how developers use these messages. Our findings show that developers only sometimes include an explanation message in their assertion method. An analysis of these messages shows that they are either composed of string literals, identifiers, or a combination of strings and identifiers, with most messages comprising of string literals. Furthermore, as part of our analysis, we report on these messages' readability and grammatical structure. Finally, we also discuss assert message anti-patterns that we discovered as part of our qualitative analysis. The discovery of these anti-patterns is an important step forward in improving the quality of assert messages.


Our future work in this area includes a human subject study. In this proposed study, we will work with developers of varying experience and skills to validate our empirical findings and gain further insight into heuristics we can incorporate into appraising and recommending high-quality assertion messages.

    

\bibliographystyle{ieeetr}
\bibliography{main}

\end{document}